\documentclass[reprint,amsmath,amssymb,aps,prb,superscriptaddress]{revtex4-2}
\usepackage{graphicx}
\usepackage{dcolumn}
\usepackage{bm}
\usepackage[mathlines]{lineno}
\usepackage[utf8]{inputenc}
\usepackage[T1]{fontenc}
\usepackage{booktabs, array, float, tabularx, lipsum, amsmath, amsthm, amssymb, multirow, esvect}
\usepackage{paralist}

\usepackage{newtxtext,newtxmath}
\DeclareSymbolFont{orilargesymbols}{OMX}{cmex}{m}{n}
\DeclareMathSymbol{\orisum}{\mathop}{orilargesymbols}{"50}
\renewcommand\sum\orisum
\renewcommand\geq\geqslant 
\renewcommand\leq\leqslant
\usepackage[cal=cm]{mathalpha}  
\usepackage{makecell}
\usepackage{siunitx}
\usepackage[dvipsnames]{xcolor}
\usepackage[version=4]{mhchem}
\usepackage[colorlinks,linkcolor=blue,anchorcolor=blue,citecolor=blue,urlcolor=blue]{hyperref}
\usepackage{orcidlink}

\begin{document}

\title{Strain-induced nonrelativistic altermagnetic spin splitting effect}

\newcommand{\WUSTa}{\affiliation{State Key Laboratory of Advanced Refractories, \href{https://ror.org/00e4hrk88}{Wuhan University of Science and Technology}, Wuhan 430081, People's Republic of China}}
\newcommand{\WUSTb}{\affiliation{School of Materials and Metallurgy, \href{https://ror.org/00e4hrk88}{Wuhan University of Science and Technology}, Wuhan 430081, People's Republic of China}}
\newcommand{\WHU}{\affiliation{Key Laboratory of Artificial Micro- and Nano-structures of Ministry of Education,\\ School of Physics and Technology, \href{https://ror.org/033vjfk17}{Wuhan University}, Wuhan 430072, People's Republic of China}}

\author{Wancheng Zhang~\orcidlink{0000-0003-1382-9806}}
\WUSTa
\WUSTb
\author{Mingkun Zheng}
\WUSTa
\WUSTb
\author{Yong Liu}
\WHU
\author{Zhenhua Zhang~\orcidlink{0000-0003-2148-8239}}
\email[Contact author:]{zzhua@wust.edu.cn}
\WUSTa
\WUSTb
\author{Rui Xiong~\orcidlink{0000-0003-0468-6014}}
\email[Contact author:]{xiongrui@whu.edu.cn}
\WHU
\author{Zhihong Lu~\orcidlink{0000-0002-6636-0581}}
\email[Contact author:]{zludavid@live.com}
\WUSTa
\WUSTb


\begin{abstract}
Recent studies reveal that $\mathcal{T}$-odd spin currents generated via the nonrelativistic altermagnetic spin splitting effect (ASSE) exhibit significant potential for spintronics applications, with both computational and experimental validations. Addressing the scarcity of conductive altermagnets, we propose strain engineering as a reliable method for inducing altermagnetism. Focusing on rutile-structured $\mathrm{OsO}_2$, first-principles calculations show that minor equibiaxial tensile strain ($\mathcal{E}_{\mathrm{ts}}$=3\%) induces nonmagnetic-to-altermagnetic transitions, achieving an ASSE-driven spin-charge conversion ratio ($\theta_{\text{AS}}$) of $\sim$7\%---far surpassing conventional spin Hall angles ($\theta_{\text{IS}}$). Calculations reveal that substantial $\theta_{\text{AS}}$ persists even in the absence of spin-orbit coupling, with its magnitude positively correlating to nonrelativistic spin splitting magnitude, which further confirms the strain-induced ASSE's nonrelativistic origin. Further investigation reveals that $\mathrm{RuO}_2$ exhibits analogous phenomena, which may resolve recent controversies regarding its magnetic properties. Our research opens new simple pathways for developing next-generation altermagnetic spintronic devices.
\begin{center}
    \begin{minipage}{\textwidth}
        \centering
        \small Published in \textbf{\textit{Physical Review B}}: \href{https://doi.org/10.1103/8zlt-mlms}{Phys. Rev. B 112, 024415 (2025)}
    \end{minipage}
\end{center}
\end{abstract}

\maketitle

\setlength {\parskip} {0pt}
\section{Introduction}\label{intro}
Altermagnetism, recently established as the third fundamental class of magnetism alongside ferromagnetism and antiferromagnetism, exhibits a unique duality: vanishing net magnetization in real space (like antiferromagnetism) combined with broken time-reversal symmetry ($\mathcal{T}$) in reciprocal space (akin to ferromagnetism). Unlike conventional antiferromagnets where opposite spin sublattices are linked by translation or inversion symmetry, altermagnets feature sublattices connected via rotational symmetry operations (proper/improper, symmorphic/non-symmorphic)~\cite{altermagnetic1,altermagnetic2,altermagnetic3,altermagnetic4}. This distinct symmetry landscape generates a nonrelativistic spin-splitting effect (analogous to even-parity $d$-, $g$-, or $i$-wave symmetry) with momentum-dependent alternating spin patterns, enabling phenomena such as the altermagnetic spin-splitting effect (ASSE), which generate a transverse pure spin current when a charge current flows along specific crystallographic directions-even in the absence of relativistic spin-orbit coupling (SOC)~\cite{SST1,SST2,RuO2-ASSE}. These spin currents arise from spin-momentum locking in the electronic structure, providing a dissipationless pathway for spin-charge interconversion, which is critical for field-free magnetization switching in magnetic random-access memory (MRAM)~\cite{rut3,rut4}.

At present, most identified altermagnetic materials are semiconductors or insulators, such as MnTe~\cite{MnTe1,MnTe2,MnTe3} and recently proposed two-dimensional altermagnetic systems fabricated by stacking and twisting~\cite{stacking1,stacking2,stacking3}. Metallic altermagnets remain scarce, with $d$-wave $\mathrm{RuO}_2$~\cite{SST1,SST2,RuO2-ASSE,RuO2-1,RuO2-2,RuO2-3,RuO2-4-F,RuO2-3-T} and $g$-wave CrSb~\cite{CrSb1,CrSb2,CrSb3} being prominent examples. Moreover, the entire spin conductivity tensors of the $^26/^2m^2m^1m$ spin Laue group with $g$-wave symmetry are forced to vanish. In this scenario, spin current generation requires the introduction of shear strain in the $xy$-plane to reconfigure its spin-momentum pattern into $d$-wave symmetry~\cite{Hex}. Although $\mathrm{RuO}_2$ has been proposed as a prototypical $d$-wave altermagnet, experimental validation remains contentious: the angle resolved photoemission spectroscopy (ARPES) and spin-resolved ARPES (SARPES) studies report no detectable momentum-dependent spin splitting~\cite{RuO2-4-F}, whereas magnetic circular dichroism (MCD) measurement unambiguously demonstrates $\mathcal{T}$-symmetry breaking in its band structure~\cite{RuO2-3-T}. The latest muon spin relaxation/rotation ($\mu$SR) studies even point to the nonmagnetism of $\mathrm{RuO}_2$~\cite{RuO2-NM1,RuO2-NM2}. This discrepancy underscores the need for alternative material platforms with unambiguous altermagnetic signatures. Here, we propose a strategy for designing altermagnetic materials: inducing altermagnetism via strain engineering in material systems with altermagnetic crystal symmetry.

In this work, we systematically investigate the strain-induced ASSE in bulk $\mathrm{OsO}_2$ using \textit{ab~initio} calculations. Bulk $\mathrm{OsO}_2$ single crystals exhibit nonmagnetic metallic behavior with a room-temperature resistivity of $\sim6\times10^{-5}~\Omega\cdot\mathrm{cm}$~\cite{lattice1,OsO2}, while the monolayer $1T$~\cite{1T-OsO2} and $1T'$~\cite{1T'-OsO2} phases of $\mathrm{OsO}_2$ are calculated to be ferromagnetic metal and nonmagnetic semiconductor, respectively. We propose equibiaxial tensile strain $\mathcal{E}_{\mathrm{ts}}$ as an effective and easily achievable method to induce altermagnetism in bulk $\mathrm{OsO}_2$. Our calculations reveal a strain-dependent alternating Fermi surface, which becomes increasingly pronounced with increasing $\mathcal{E}_{\mathrm{ts}}$. In the absence of Hubbard $U$ correction, this alternating pattern is observed at the $k_{z}=\pi/2c$ plane and disappears when $\mathcal{E}_{\mathrm{ts}}$ reaches $6\%$. Using maximally localized Wannier functions in conjunction with linear response theory and the Kubo formula (detailed in Sec.~\ref{Math}), we calculate the ASSE-induced $\mathcal{T}$-odd spin conductivity $\sigma_{xy}^{z\text{,odd}}$ and spin-charge conversion ratio $\theta_{\text{AS}}$. Enhancement of both the nonrelativistic altermagnetic spin-splitting conductivity (ASSC) and $\theta_{\text{AS}}$ with an increase in $\mathcal{E}_{\mathrm{ts}}$ is observed. These values are found to be significantly larger than those of the conventional SOC-dependent $\mathcal{T}$-even intrinsic spin Hall conductivity (ISHC) $\sigma_{xy}^{z}$ and its associated spin-charge conversion ratio (i.e., spin Hall angle) $\theta_{\text{IS}}$.

Our study establishes a feasible method for strain-induced altermagnetism, and identifies $\mathrm{OsO}_2$ as a potential altermagnet and viable platform for field-free switching of perpendicular magnetization in MRAM devices. The ASSE-generated spin currents circumvent the limitations of relativistic conventional spin Hall effect (CSHE) mechanisms, offering enhanced efficiency and scalability compared to conventional spin-orbit torque (SOT) mechanisms. By bridging the gap between altermagnetic theory and functional material design, this work opens new avenues for materials exhibiting altermagnetic crystal symmetry in next-generation spintronics.

\section{Methodology}\label{Math}
Our first-principles calculations are implemented in the Vienna $ab~initio$ Simulation Package (\textsc{vasp}) following the density functional theory~\cite{VASP1,VASP2,VASP3}. Projected augmented-wave pseudopotentials are utilized to describe the ion-electron interaction, while Perdew-Burke-Ernzerhof (PBE) and the generalized gradient approximation (GGA) are adopted as the exchange-correlation potentials~\cite{PAW1994,GGA1996}. Besides, the cutoff energy of the plane-wave basis is set to 520 eV. In the process of structure optimization, the convergence criteria of energy and residual force are set to $1\times10^{-8}$ eV and 0.01 eV/\AA, respectively. During structural optimization, the Brillouin zone (BZ) is sampled using a $\Gamma$-centered $7\,\times\,7\,\times\,11$ Monkhorst-pack $k$ mesh. In self-consistent field (SCF) calculations, a $9\,\times\,9\,\times\,13$ Monkhorst-pack $k$ mesh is employed to obtain an accurate electronic structure. In this work, maximally localized Wannier functions and the Kubo formula are utilized to calculate the ISHC~\cite{guo-prl2008,log1}. The tight-binding models are calculated by \textsc{wannier90}~\cite{wannier90}. To handle the rapid variation of the spin Berry curvature (SBC), the BZ integration is conducted using a dense $k$ mesh with $500\times500$. 

Since the response of metals to electric fields can be well described by the linear response theory, the $\mathcal{T}$-odd spin conductivity and spin-charge conversion ratio within the linear response theory are evaluated using the Kubo formula in the approximation of the constant scattering rate $\bm{\Gamma}$, as implemented in the \textsc{wannier-linear-response} code~\cite{SST1,SST2,WLR1}. In this constant $\bm{\Gamma}$ approximation, it is assumed that the only effect of disorder is a constant band broadening, which modifies Green's functions of the perfect periodic system in the following approach: $G^{R}(\varepsilon)=1 /(\varepsilon-\hat{H}+i 0+) $ $\to$ $ 1 /(\varepsilon-\hat{H}+i \bm{\Gamma})$, where $\hat{H}$ denotes the Hamiltonian, $\varepsilon$ denotes energy, and $G^{R}$ is the retarded Green's function~\cite{PhysRevB.77.165117}. The Kubo formula within the constant $\bm{\Gamma}$ approximation can be split into the $\mathcal{T}$-odd contribution~\cite{T-odd}
\begin{widetext}
\begin{equation}
    \label{eq:T_odd}
    \sigma_{\alpha\beta}^{\gamma\text{,odd}} = -\frac{e\hbar}{\pi}\int \frac{d^3\bm{k}}{(2\pi)^3}\sum_{ n,m} \frac{\bm{\Gamma}^2 \text{Re}
    (\langle \psi_{n\bm{k}}| \hat{A}|\psi_{m\bm{k}}\rangle  \langle \psi_{m\bm{k}}| \hat{v}_\beta | \psi_{n\bm{k}}\rangle )}{[(E_F-\varepsilon_{n\bm{k}})^2+\bm{\Gamma}^2]
    [(E_F-\varepsilon_{m\bm{k}})^2+\bm{\Gamma}^2]}\text{,} 
\end{equation}
\end{widetext}
and the $\mathcal{T}$-even contribution given in the $\bm{\Gamma}~\to~0$ limit by~\cite{T-even}
\begin{widetext}
\begin{equation}
    \label{eq:T_even}
    \sigma_{\alpha\beta}^{\gamma\text{,even}}=-2e\hbar\int\frac{d^3\bm{k}}{(2\pi)^3}\sum_{n\neq m}^{\substack{n~\text{occ}\\m~\text{unocc}}}\frac{\mathrm{Im}
    (\langle \psi_{n\bm{k}}|\hat{A}|\psi_{m\bm{k}}\rangle  \langle \psi_{m\bm{k}}|\hat{v}_\beta|\psi_{n\bm{k}}\rangle)}{(\varepsilon_{n\bm{k}}-\varepsilon_{m\bm{k}})^2}\text{,}
\end{equation}
\end{widetext}
where $e$ represents the (positive) elementary charge; $\alpha$, $\beta$, $\gamma$ $=$ $x$, $y$, $z$ represent the directions of spin current, electric field, and spin polarization, respectively; 
$\bm{k}$ stands for the Bloch wave vector; $n$ and $m$ are the band indices; $\psi_{n\bm{k}}$, $\varepsilon_{n\bm{k}}$ denote the Bloch function for band $n$ at $\bm{k}$ and the corresponding band energy, respectively; 
$E_F$ denotes the Fermi energy, and $\hat{v}_\beta$ is the velocity operator. In Eq.~\eqref{eq:T_even}, the sum is restricted to $m$, $n$ such that $n$ is occupied and $m$ is unoccupied. 
Eqs.~\eqref{eq:T_odd},~\eqref{eq:T_even} can describe the spin conductivity by setting operator $\hat{A}=\hat{j}_{\alpha}^{\gamma}$, where $\hat{j}_{\alpha}^{\gamma}=\frac{1}{2}\{\hat{s}_{\gamma}$,$\hat{v}_{\alpha}\}$ is the spin current operator and $\hat{s}_{\gamma}=\frac{\hbar}{2}\hat{\sigma}_{\gamma}$ is the spin operator.
Eq.~\eqref{eq:T_odd} can be further adjusted to calculate the charge conductivity by simply setting the operator $\hat{A}=-e\hat{v}_{\alpha}$, while changing the left side of the equation to $\sigma_{\alpha\beta}$~\cite{WLR1,T-odd,CrXO}. A constant $\bm{\Gamma}$ that determines the broadening magnitude is used, which can be estimated by comparing the calculated conductivity with the experimental conductivity. 
Time reversal is an antiunitary operator that will transform the matrix elements as 
$\langle \psi_{n\bm{k}}|\hat{A} | \psi_{n\bm{k}} \rangle$ $\to$ $\langle \psi_{n\bm{k}}|\mathcal{T} \hat{A} \mathcal{T}| \psi_{n\bm{k}}\rangle^{*}$~\cite{PhysRevB.95.014403}, so Eqs.~\eqref{eq:T_odd} and \eqref{eq:T_even} will transform differently under time reversal. It should be noted that the transformation under time reversal is the opposite for conductivity and spin conductivity. This is because the spin current operator contains an additional spin operator that is odd under time reversal. Thus, for spin conductivity, Eq.~\eqref{eq:T_odd} is odd under time reversal, while Eq.~\eqref{eq:T_even} is even~\cite{WLR1}.

For the convenience of SBC's calculation, Eq.~\eqref{eq:T_even} can be further rewritten into the form of Eq.~\eqref{eq:kubo_shc}, which is also known as the Kubo-Greenwood formula for the direct current (DC) ISHC of a crystal in the independent-particle approximation, and it is written as ~\cite{log1,guo-prl2008}
\begin{widetext}
\begin{equation}
  \label{eq:kubo_shc}
  \sigma_{\alpha\beta}^{\gamma} =  \frac{\hbar}{\mathcal{V} _c \mathcal{N}_{\bm{k}}} \sum_{\bm{k}}\sum_{n} f_{n\bm{k}} \sum_{m \neq n}\frac{2\operatorname{Im}[\langle \psi_{n\bm{k}}| \hat{j}_{\alpha}^{\gamma}|\psi_{m\bm{k}}\rangle \langle \psi_{m\bm{k}}| -e\hat{v}_{\beta}|\psi_{n\bm{k}}\rangle]}{(\varepsilon_{n\bm{k}}-\varepsilon_{m\bm{k}})^2}\text{,}
\end{equation}
\end{widetext}
where $\mathcal{V}_c$ represents the cell volume, $\mathcal{N}_{\bm{k}}$ represents the number of $k$-points used for sampling the Brillouin zone, and $f_{n{\bm k}}=f(\varepsilon_{n{\bm k}})$ represents the Fermi-Dirac distribution function.

Eq.~(\ref{eq:kubo_shc}) can be further divided into the
band-projected Berry curvature-like term 
\begin{equation}
  \label{eq:kubo_shc_berry}
  \!\Omega_{n,\alpha\beta}^{\gamma}(\bm{k})\! =\! {\hbar}^2 \!\sum_{
  m\ne n}\!\frac{-2\!\operatorname{Im}[\!\langle \psi_{n\bm{k}\!} \left|
  \frac{1}{2}\!\{\hat{\sigma}_{\gamma},\hat{v}_{\alpha}\}\right|\!\psi_{m\bm{k}}\rangle\!
  \langle \psi_{m\bm{k}\!}\left|\hat{v}_{\beta}\right|\psi_{n\bm{k}}\rangle\!]}
  {(\varepsilon_{n\bm{k}}-\varepsilon_{m\bm{k}})^2}\text{,}
\end{equation}
and the $k$-resolved term that sums over occupied bands
\begin{equation}
  \label{eq:kubo_shc_berry_sum}
  \Omega_{\alpha\beta}^{\gamma}(\bm{k}) = \sum_{n}
  f_{n\bm{k}} \Omega_{n,\alpha\beta}^{\gamma}(\bm{k})\text{,}
\end{equation}
and the SHC can be represented as
\begin{equation}
  \label{eq:shc}
  \sigma_{\alpha\beta}^{\gamma} =
  \frac{e}{2}\frac{1}{\mathcal{V} _c \mathcal{N}_{\bm{k}}}\sum_{\bm{k}}
  \Omega_{\alpha\beta}^{\gamma}(\bm{k})\text{.}
\end{equation}
The unit of $\Omega_{n,\alpha\beta}^{\gamma}(\bm{k})$ is $\mathrm{length}^2$, 
and the unit of  $\sigma_{\alpha\beta}^{\gamma}$ 
is $(\hbar/e)\mathrm{S}/\mathrm{length}$.

\section{Discussion}\label{RD}
\subsection{\label{basis}Crystal and electronic band structure of OsO$\mathbf{_2}$}

\begin{figure}
  \centering
  \includegraphics[width=0.5\textwidth]{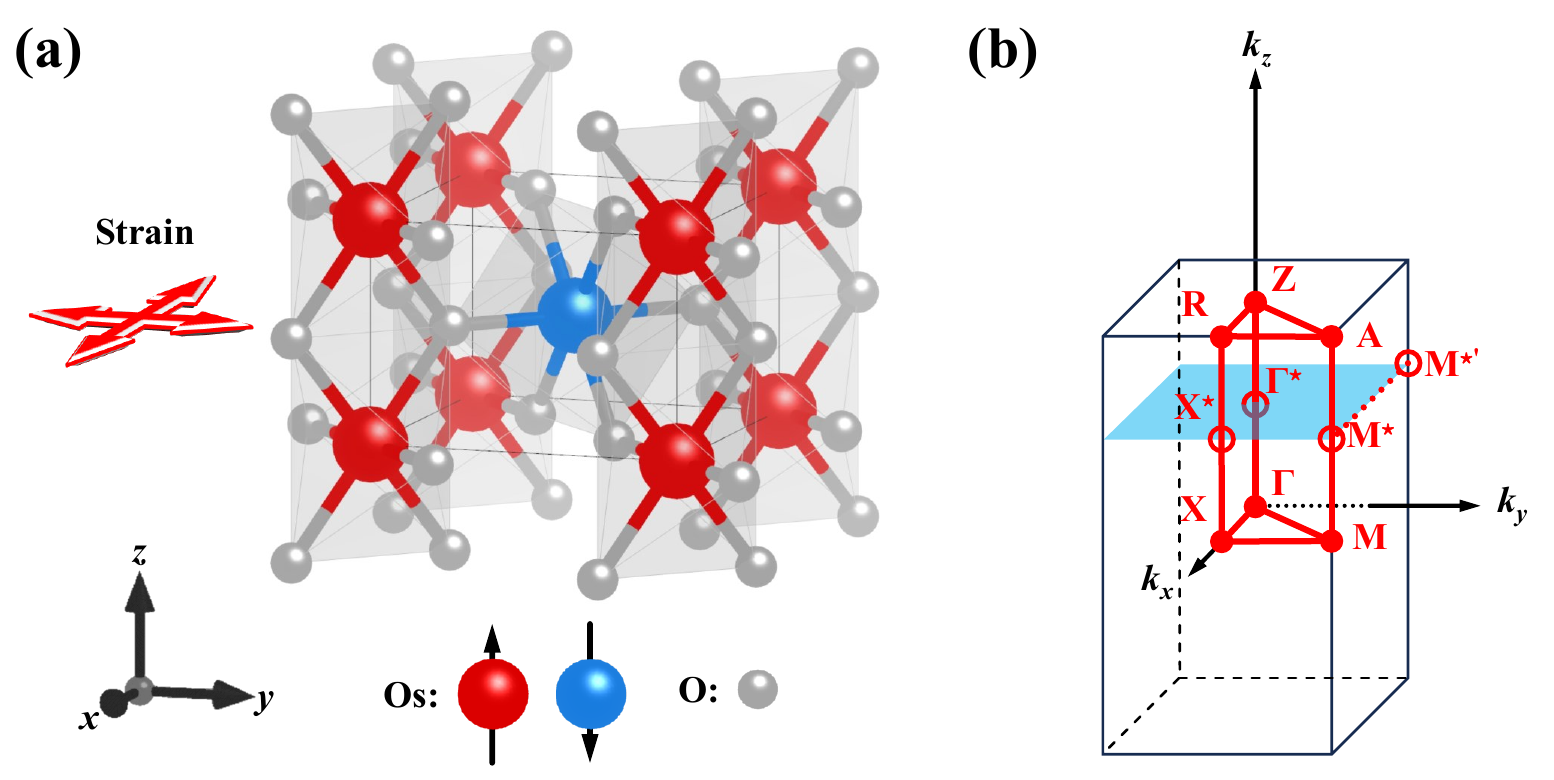}
  \caption{\label{Structure}(a) The crystal structure of bulk rutile $\mathrm{OsO}_2$. The red arrow indicates the direction of equibiaxial tensile strain $\mathcal{E}_{\mathrm{ts}}$; the red and blue spheres represent Os atoms with spin-up and spin-down states respectively, and the gray spheres represent oxygen atoms. (b) The schematic diagram of $\mathrm{OsO}_2$'s three-dimensional Brillouin zone (BZ), with high-symmetry points indicated by red dots. The blue cross-section indicates the $k_{z}=\pi/2c$ plane where the alternating pattern occurs.
  }
\end{figure}

$\mathrm{OsO}_2$ crystallizes in the well-known rutile structure~\cite{rut1,rut2} [as shown in Fig.~\hyperref[Structure]{\ref{Structure}(a)}], with experimental lattice constants of $a \approx 4.50$ \AA~and $c \approx 3.18$ \AA~\cite{lattice1,lattice2,lattice3}. This is close to our structural optimization results ($a = 4.522$ \AA~and $c = 3.215$ \AA) without considering strain. 
The results from Ref.~\cite{OsO2-DFT} demonstrate that the DFT calculations (partial density of states) of $\mathrm{OsO}_2$ without considering the Hubbard $U$ correction agree well with the experiment (high-resolution valence band spectra). Therefore, our discussion in the main text focuses mainly on the $U=0$ case. Simultaneously, we address the scenario with on-site Coulomb interaction in \hyperref[sec:appendix]{Appendix}.
Unlike $\mathrm{RuO}_2$, where experimental data support the selection of $U$ values, current research lacks experimental studies on the magnetism of $\mathrm{OsO}_2$. Consequently, we determined $U = 1.13$ eV via the linear response approach~\cite{lla-U}, as shown in Fig.~\ref{FLLAU}. Notably, at this $U$ value, $\mathrm{OsO}_2$ remains non-magnetic. Additionally, we also examined the evolution of magnetism in $\mathrm{OsO}_2$ under different $U$ values (see Table~S1 in Supplemental Material~\cite{supplement}). Notably, when the Coulomb interaction term $U \geqslant 1.5$ eV, $\mathrm{OsO}_2$ exhibits more pronounced altermagnetism compared to the strain-induced effects observed in the absence of $U$. For the sake of rigor, we selected $U = 2.0$ eV to simulate the scenario where $\mathrm{OsO}_2$ intrinsically possesses altermagnetism, and we examined strain effects on the ASSE-driven spin-charge conversion ratio ($\theta_{\text{AS}}$) under this condition (Table~S2~\cite{supplement}). This comprehensive approach enables systematic investigation of strain effects on $\mathrm{OsO}_2$, whose magnetic ground state is experimentally undetermined.

When the on-site Coulomb interaction was not taken into account, we conducted tests on equibiaxial strain ($xy$ plane) of $\mathrm{OsO}_2$ at intervals of $1\%$ within the range from $-6\%$ to $+6\%$. We found that only when the strain is greater than $2\%$ (i.e., equibiaxial tensile strain $\mathcal{E}_{\mathrm{ts}}$), does $\mathrm{OsO}_2$ exhibit relatively obvious altermagnetism. We also noticed that when $\mathcal{E}_{\mathrm{ts}}$ is within the range of $2\% \sim 5\%$, its magnetism is enhanced as $\mathcal{E}_{\mathrm{ts}}$ increases. However, when $\mathcal{E}_{\mathrm{ts}}$ reaches $6\%$, its magnetism instead weakens. Since equibiaxial compressive strain cannot induce magnetism in $\mathrm{OsO}_2$, this work only considers the case of equibiaxial tensile strain ($\mathcal{E}_{\mathrm{ts}}$). We list the changes of the lattice constants of $\mathrm{OsO}_2$, the magnetic moment of Os atoms, and the maximum splitting in the bands of $\mathrm{OsO}_2$ near the Fermi level with $\mathcal{E}_{\mathrm{ts}}$ ranging from $0\%$ to $6\%$ in Table~\ref{tab1}.

\begin{table*}
  \caption{\label{tab1}Evolution of $\mathrm{OsO}_2$'s lattice constants, magnetic moments, spin splitting, and spin-resolved Fermi surface shapes under different $\mathcal{E}_{\mathrm{ts}}$.}
  \tabcolsep=0.48cm
  \begin{ruledtabular}
    \begin{tabular}{cccccc}
      \multirow{2}{*}{$\mathcal{E}_{\mathrm{ts}}~(\%)$} & \multicolumn{2}{c}{Lattice constants (\AA)} & \multirow{2}{*}{\makecell[c]{Magnetic moment \\ of Os atoms ($\mu_{B}$)}} & \multirow{2}{*}{\makecell[c]{$\left | \mathrm{Splitting} \right |_\mathrm{max}$ (meV)\\ near the Fermi level}} & \multirow{2}{*}{\makecell[c]{Spin-resolved Fermi surface\\ $@$ $k_{z}=\pi/2c$ plane (with SOC)}}   \\ \cline{2-3}  
                                    & $a$          & $c$          &                        &                        &                                                                 \\ [3pt] \hline & \\[-2.0ex]
        0                           & $4.522$      & $3.215$      & $\pm$ $0.000$          & $-       $             & $\raisebox{-.5\height}{\includegraphics[height=1cm]{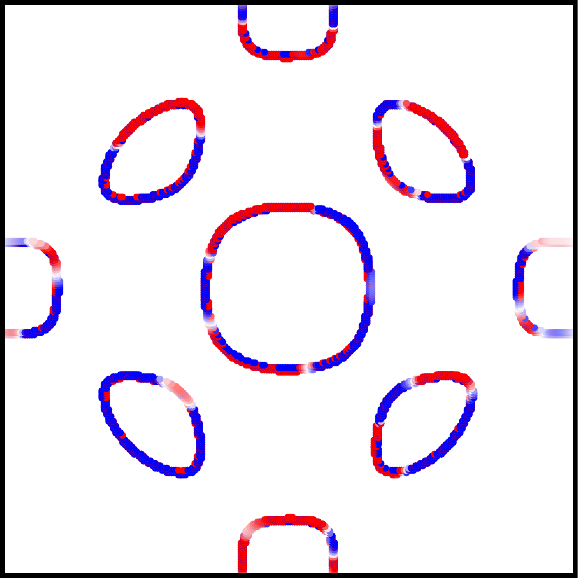}}$ \\ [15pt]
        1                           & $4.568$      & $3.169$      & $\pm$ $0.002$          & $62.8~@  $ $A-Z$       & $\raisebox{-.5\height}{\includegraphics[height=1cm]{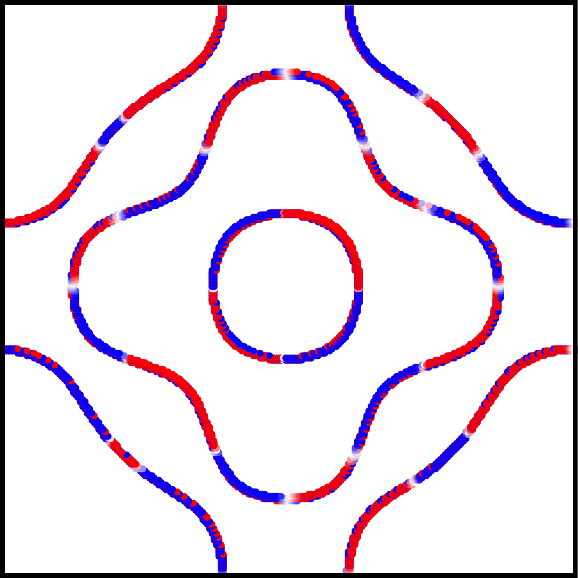}}$ \\ [15pt]
        2                           & $4.612$      & $3.125$      & $\pm$ $0.028$          & $209.2~@ $ $\Gamma-M$  & $\raisebox{-.5\height}{\includegraphics[height=1cm]{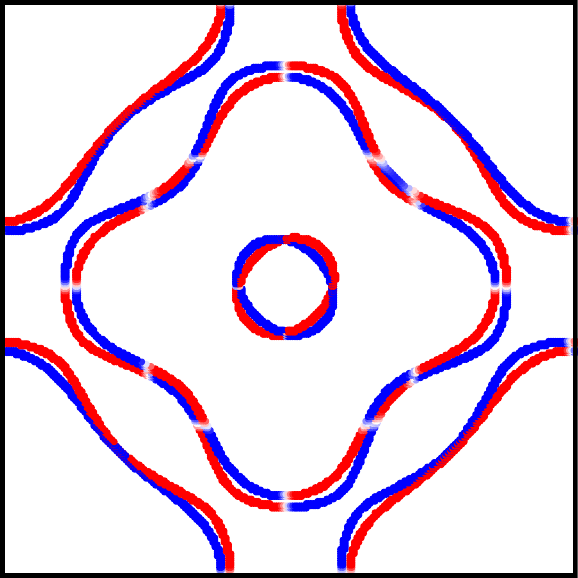}}$ \\ [15pt]
        3                           & $4.658$      & $3.080$      & $\pm$ $0.349$          & $256.3~@ $ $\Gamma-M$  & $\raisebox{-.5\height}{\includegraphics[height=1cm]{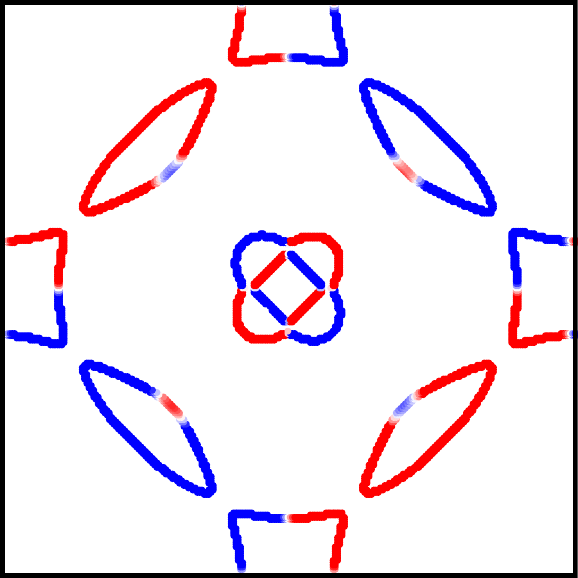}}$ \\ [15pt]
        4                           & $4.703$      & $3.043$      & $\pm$ $0.468$          & $275.5~@ $ $A-Z$       & $\raisebox{-.5\height}{\includegraphics[height=1cm]{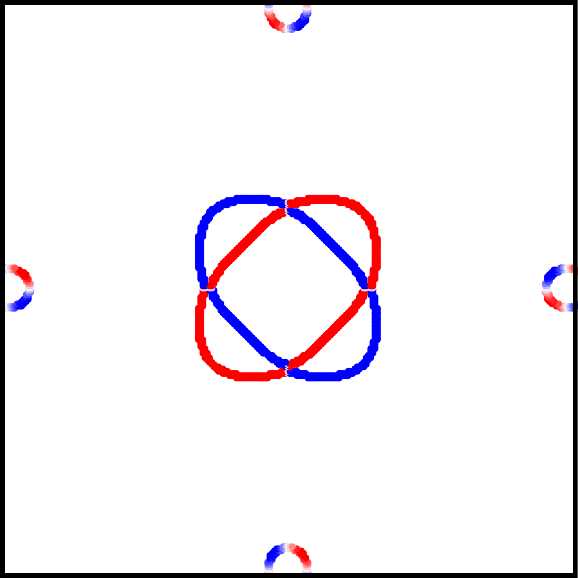}}$ \\ [15pt]
        5                           & $4.749$      & $3.003$      & $\pm$ $0.500$          & $110.0~@ $ $A-Z$       & $\raisebox{-.5\height}{\includegraphics[height=1cm]{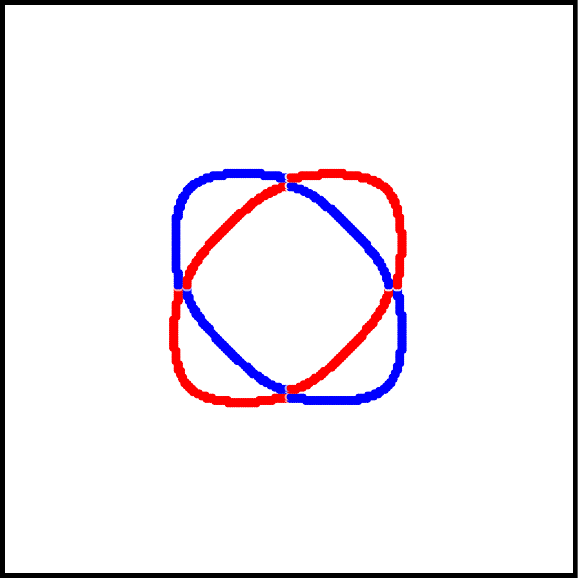}}$ \\ [15pt]
        6                           & $4.795$      & $2.965$      & $\pm$ $0.150$          & $0.11~@  $ $A-Z$       & $\raisebox{-.5\height}{\includegraphics[height=1cm]{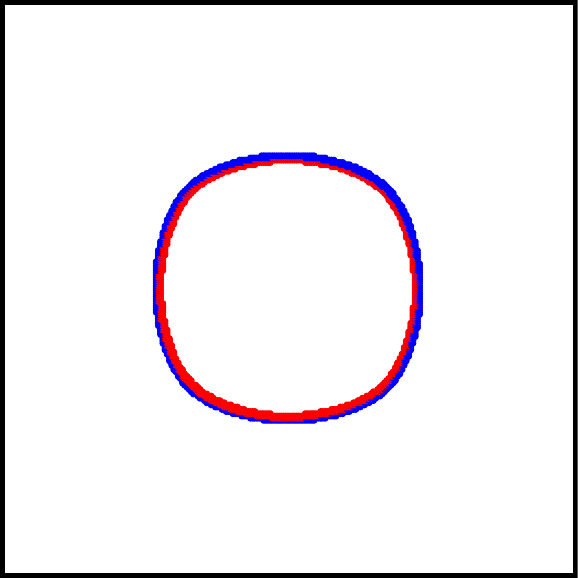}}$                   
    \end{tabular}
  \end{ruledtabular}
\end{table*}

The electronic band structures and density of states (DOS) of $\mathrm{OsO}_2$ under different $\mathcal{E}_{\mathrm{ts}}$ without SOC are shown in Fig.~S1~\cite{supplement}. The metallicity remains unchanged with the variation of $\mathcal{E}_{\mathrm{ts}}$ from $1\%$ to $6\%$. When $\mathcal{E}_{\mathrm{ts}}$ reaches $2\%$, a subtle spin splitting can be observed. As $\mathcal{E}_{\mathrm{ts}}$ increases to $6\%$, a tendency towards degeneracy emerges. When $\mathcal{E}_{\mathrm{ts}}$ is equal to $2\%$ and $3\%$, the maximum splitting occurs along the $\Gamma$ - $M$ path. In other cases, the maximum splitting occurs along the $A$ - $Z$ path. We provide a clearer illustration of the spin splitting and its magnitude in Fig.~S2~\cite{supplement}, and the orbital-resolved SOC band structures are displayed in Fig.~S3~\cite{supplement}.

As discussed previously, it can be observed that both the magnetic moment of Os atoms and the maximum splitting exhibit a trend of first increasing and then decreasing as $\mathcal{E}_{\mathrm{ts}}$ increases. We know that the equibiaxial tensile strain in the $xy$ axis direction (an increase in $a$ and $b$) will inevitably lead to the compressive strain towards the center of the material along the $z$ axis (a decrease in $c$) due to the Poisson effect~\cite{Poisson}. This is also consistent with our calculation results, as shown in Table~\ref{tab1}, where we can observe that the decrease in the lattice constant $c$ is slightly greater than the increase in $a$.

\subsection{\label{FermiSurface}Strain-induced special alternating Fermi surface}

\begin{figure*}
  \includegraphics[width=1.0\textwidth]{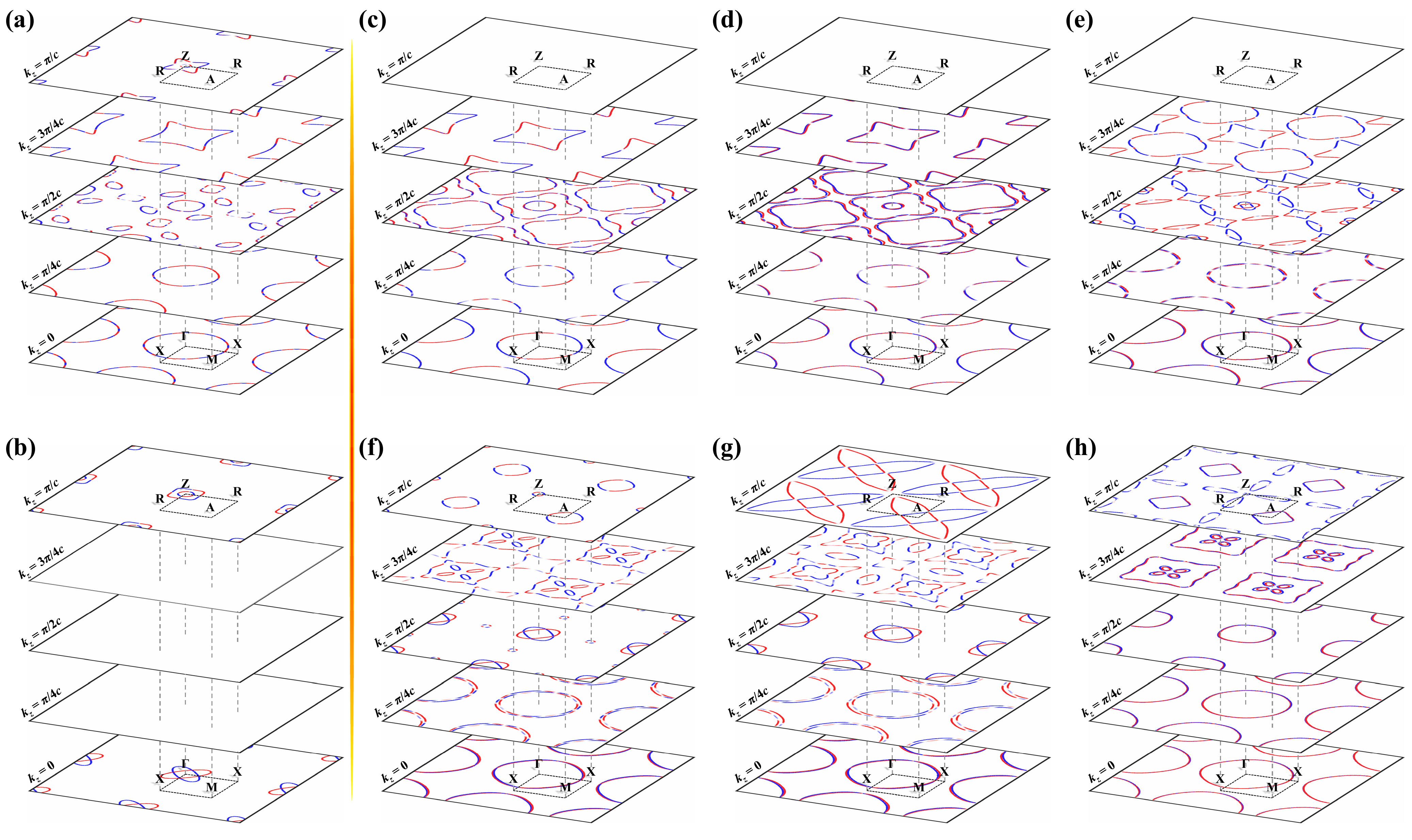}
  \caption{\label{FS}The shapes of spin-resolved Fermi surfaces at different $k_{z}$ planes for $\mathrm{OsO}_2$ under $\mathcal{E}_{\mathrm{ts}}$ values of (a) $0\%$, (b) $0\%$ ($U=2.0$ eV), (c) $1\%$, (d) $2\%$, (e) $3\%$, (f) $4\%$, (g) $5\%$, and (h) $6\%$. The red and blue colors represent spin-up and spin-down, respectively.
  }
\end{figure*}

It is noteworthy that when $\mathcal{E}_{\mathrm{ts}}$ is within the range of $2\% \sim 5\%$, the spin-resolved Fermi surface of $\mathrm{OsO}_2$ at the $k_{z}=\pi/2c$ plane shows an obvious alternating pattern, which is also listed in Table~\ref{tab1}. This is strong evidence for the altermagnetism of $\mathrm{OsO}_2$. We also calculated the 3D spin-resolved Fermi surface of $\mathrm{OsO}_2$ under different $\mathcal{E}_{\mathrm{ts}}$ (see Fig.~S4 in the Supplemental Material~\cite{supplement}). However, due to the complex shape, it is difficult to observe the internal situation through the outer surface from a 3D perspective. Therefore, we presented the projections of the 3D spin-resolved Fermi surface of $\mathrm{OsO}_2$ at different $k_{z}$ planes in the form of slices, as shown in Fig.~\ref{FS}. Interestingly, for the case without strain, even in the absence of magnetism, the spin-resolved Fermi surface of $\mathrm{OsO}_2$ still exhibits an alternating pattern at the $k_{z}=\pi/c$ plane, as shown in Fig.~\hyperref[FS]{\ref{FS}(a)} and Fig.~S5~\cite{supplement}. As the slices move downward, the spin-up and spin-down components (i.e., the red and blue parts) gradually blend together and become indistinguishable, which can be observed more clearly in Fig.~S5~\cite{supplement}. At present, we are not clear about the specific reasons for this phenomenon, and it is also beyond the scope of discussion in this paper. However, its physical origin is still worthy of further investigation. Fig.~\hyperref[FS]{\ref{FS}(b)} shows the spin-resolved Fermi surface of $\mathrm{OsO}_2$ with $\mathcal{E}_{\mathrm{ts}}=0\%$ and Hubbard parameter $U=2$ eV for comparison. The alternating pattern can be observed at the $k_{z} = 0 $ and $k_{z}=\pi/c$ planes, while the other slices are very ``clean'', which is similar to the results of DFT + $U$ calculations for $\mathrm{RuO}_2$ (the 3D spin-resolved Fermi surface of $\mathrm{RuO}_2$ is displayed in Fig.~S4~\cite{supplement}).

\begin{figure*}
  \includegraphics[width=1.0\textwidth]{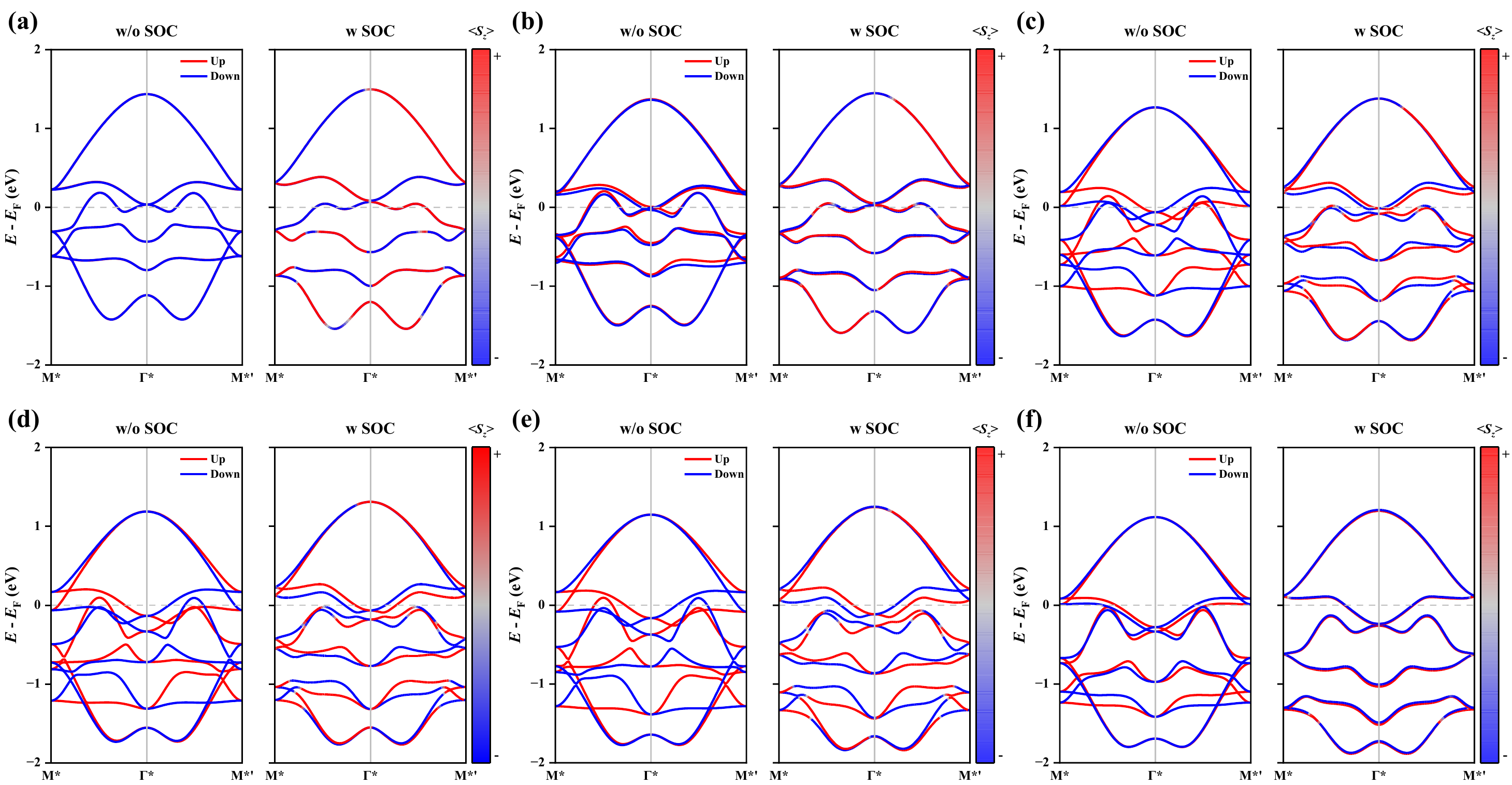}
  \caption{\label{025band}Evolution of band structures with $\mathcal{E}_{\mathrm{ts}}$ ranging from $1\%$ to $6\%$ [labeled (a) to (f)] at the $k_{z}=\pi/2c$ plane with and without SOC. The positions of the high symmetry points $\Gamma^{\star}$, $M^{\star}$, and $M^{\star}$$'$ are displayed in Fig.~\hyperref[Structure]{\ref{Structure}(b)}. Red and blue represent opposite spins.
  }
\end{figure*}

As can be observed from Table~\ref{tab1} and Fig.~\ref{FS}, with the increase of $\mathcal{E}_{\mathrm{ts}}$, the spin-up and spin-down states gradually split along a specific direction under the action of the crystal field, and an alternating pattern is formed at the $k_{z}=\pi/2c$ plane, which is similar to the case in CrSb~\cite{CrSb2}, indicating the emergence of altermagnetism. Our orbital-resolved band structure analysis reveals that the electronic states near the Fermi level in $\mathrm{OsO}_2$ are predominantly contributed by the Os $d_{z^2}$, $d_{x^2-y^2}$, and $d_{xz}+d_{yz}$ orbitals. Under increasing $\mathcal{E}_{\mathrm{ts}}$, the $d_{z^2}$-dominated bands exhibit contrasting evolution trends: they shift downward along the $\Gamma$-$X$-$M$-$\Gamma$ path ($k_{z}=0$ plane) while moving upward along the $Z$-$R$-$A$-$Z$ path ($k_{z}=\pi/c$ plane), eventually crossing the Fermi level at the $k_{z}=\pi/c$ plane (Fig.~S3~\cite{supplement}). The $k_{z}=\pi/2c$ plane is exactly the momentum region lying between the $k_{z}=0$ and $k_{z}=\pi/c$ planes. It may serve as the boundary region for two opposite trends of band evolution, leading to the alternation of the spin polarization and the formation of an alternating pattern. We plot the band structures of $\mathrm{OsO}_2$ at the $k_{z}=\pi/2c$ plane with $\mathcal{E}_{\mathrm{ts}}$ increasing from $1\%$ to $6\%$, as shown in Fig.~\ref{025band}. At the Fermi level, the band evolution along the $\Gamma^{\star}$-$M^{\star}$ path can be well correlated with Fig.~\ref{FS} and Table~\ref{tab1}. These split bands that form the alternating patterns near the Fermi level are mainly contributed by Os $d_{xz}+d_{yz}$ and $d_{x^{2}-y^{2}}$ orbitals. As can be clearly observed from Fig.~\ref{025band}, even in the absence of SOC, the altermagnetic spin splitting can still occur, which further confirms its nonrelativistic origin. In addition, it is also discernible that the introduction of SOC instead weakens the spin splitting, resulting in a narrower gap between the spin-up and spin-down bands, espetially for $\mathcal{E}_{\mathrm{ts}}=4\%$ and $5\%$. This may lead to a weakening of the ASSE-induced transverse spin current, which will be discussed in detail in the next section (Sec.~\ref{SSE}).

It is noted that as $k_{z}$ becomes larger, the pattern on the $k$ slice changes more and more drastically with the variation of $\mathcal{E}_{\mathrm{ts}}$. When $\mathcal{E}_{\mathrm{ts}}$ increases from $1\%$ to $5\%$, the red and blue concentric circles at the $k_{z}=0$ plane only expand slightly and separate a little; for the $k_{z}=\pi/2c$ plane, we can clearly observe how the alternating elliptical patterns evolve; while for the $k_{z}=\pi/c$ plane, the changes are quite drastic. When $\mathcal{E}_{\mathrm{ts}}$ is further increased to $6\%$ [Fig.~\hyperref[FS]{\ref{FS}(h)}], the spin-up and spin-down states begin to degenerate, which indicates that a greater $\mathcal{E}_{\mathrm{ts}}$ may lead to the disappearance of altermagnetism. From Table~\ref{tab1}, we can clearly see the correlation between the magnetic moment of Os atoms, the magnitude of spin splitting, and the shape of the spin-resolved Fermi surface. In Sec.~\ref{SSE}, we will further explore the relationship between these factors and the ASSE-induced $\mathcal{T}$-odd spin conductivity $\sigma_{xy}^{z\text{,odd}}$, as well as its spin-charge conversion ratio $\theta_{\text{AS}}$.

\subsection{\label{SSE}nonrelativistic ASSC induced by strain}

An important application of altermagnets in the field of spintronics is the $\mathcal{T}$-odd spin current generated by the nonrelativistic ASSE. Although the magnetism of $\mathrm{RuO}_2$ is still controversial, its significant spin-charge conversion efficiency has been confirmed both in theoretical calculations~\cite{SST1,SST2} and experiments~\cite{SST2,RuO2-ASSE,RuO2-1,RuO2-2}. This new mechanism enables an applied in-plane electrical current to generate a pure spin current polarized along the N\'eel vector in the out-of-plane direction from the $\mathrm{RuO}_2$ film into a recording FM layer. This makes it possible to control the direction of spin polarization by adjusting the direction of the N\'eel vector~\cite{CrXO}. We also calculated the magnetocrystalline anisotropy energy (MAE) of $\mathrm{OsO}_2$ with $\mathcal{E}_{\mathrm{ts}}$ ranging from $2\%$ to $5\%$, as shown in Fig.~S6~\cite{supplement}. It can be observed that the N\'eel vector $\hat{\mathcal{N}}$ of $\mathrm{OsO}_2$ remains aligned along the $[001]$ direction (i.e., the $z$-axis), consistent with $\mathrm{RuO}_2$. Calculations show that in $\mathrm{RuO}_2$, the spin-charge conversion ratio of the $\mathcal{T}$-odd ASSC induced by the ASSE can reach an astonishing 28\%~\cite{SST1}, and its characteristic of not being dependent on SOC further makes ASSE a highly promising theoretical guideline for the design of next-generation spintronic devices.

\begin{figure*}
  \includegraphics[width=1.0\textwidth]{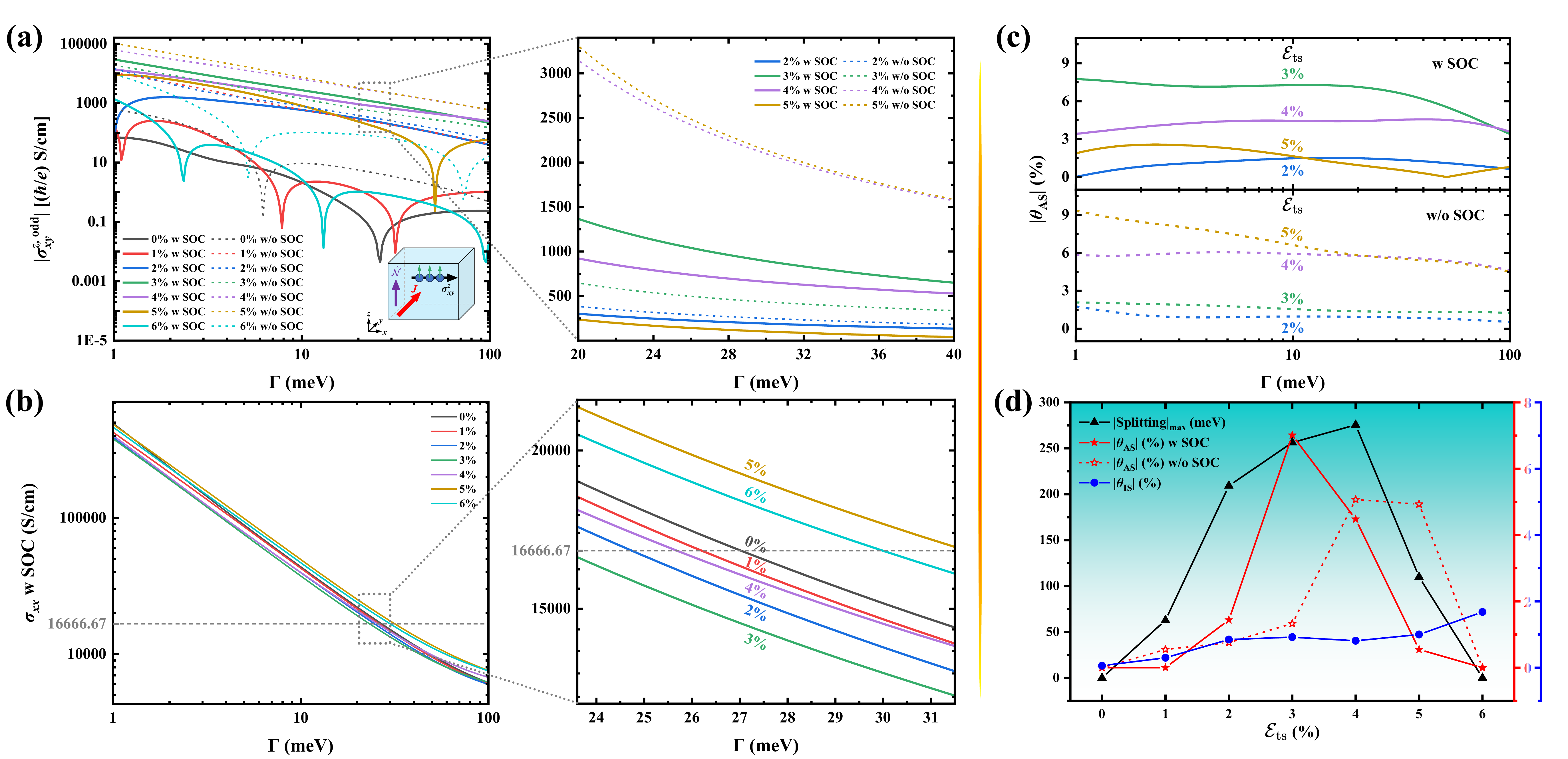}
  \caption{\label{SHC-odd}(a) $\mathcal{T}$-odd ASSC $|\sigma_{xy}^{z\text{,odd}}|$, (b) charge conductivity $\sigma_{xx}$, and (c) spin-charge conversion ratio $|\theta_{\text{AS}}|$ as functions of the scattering rate $\bm{\Gamma}$. (d) The variation of $|\mathrm{Splitting}|_{\mathrm{max}}$, $|\theta_{\text{AS}}|$, and $|\theta_{\text{IS}}|$ with $\mathcal{E}_{\mathrm{ts}}$. The inset in the lower right corner of (a) reveal the directions of charge current $J$ (red arrow), N\'eel vector (marked as $\hat{\mathcal{N}}$ with a violet arrow), spin polarization (indicated by the green arrow), and the spin current (denoted by the black arrow). The insets on the right side of (a) and (b) illustrate local details. The solid and dashed lines in (a), (c), and (d) represent results with and without SOC, respectively. The intensity of the background color in (d) indicates the magnitude of the values.
  }
\end{figure*}

In this section, we systematically investigated the strain-induced nonrelativistic ASSE in $\mathrm{OsO}_2$. Using linear response theory and Eq.~\eqref{eq:T_odd}, we calculated the $\mathcal{T}$-odd ASSC $\sigma_{xy}^{z\text{,odd}}$ and charge conductivity $\sigma_{xx}$ by setting $\hat{A}$ to $\hat{j}_{\alpha}^{\gamma}$ and $-e\hat{v}_{\alpha}$, respectively, as well as their ratio $|\theta_{\text{AS}}|=|\sigma_{xy}^{z\text{,odd}}/\sigma_{xx}|$, as shown in Fig.~\ref{SHC-odd}. Typically, we determine the value of the scattering rate $\bm{\Gamma}$ based on experimentally measured conductivity. The room-temperature conductivity of $\mathrm{OsO}_2$ is $\sim16666.67$~$\mathrm{S/cm}$~\cite{lattice1,OsO2}, as shown by the gray dashed line in Fig.~\hyperref[SHC-odd]{\ref{SHC-odd}(b)}, indicating that the corresponding $\bm{\Gamma}$ value at $\mathcal{E}_{\mathrm{ts}}=0\%$ should be $\sim27$ meV. A challenging issue is the absence of experimental conductivity data for $\mathrm{OsO}_2$ under strain, leaving no basis for determining $\bm{\Gamma}$ in such conditions. For the convenience of comparison, we assume that strain does not alter the conductivity of $\mathrm{OsO}_2$; thus, for all $\mathcal{E}_{\mathrm{ts}}$ from $0\%$ to $6\%$, we adopt the experimental value of $16666.67~\mathrm{S/cm}$ to determine $\bm{\Gamma}$. In fact, the strains involved in this work are extremely small and likely have negligible effects on conductivity. More importantly, the calculated ASSC $\sigma_{xy}^{z\text{,odd}}$ and charge conductivity $\sigma_{xx}$ exhibit an approximately linear relationship with $\bm{\Gamma}$, resulting in their ratio $|\theta_{\text{AS}}|$ (exactly what we focus on) being robust against variations in $\bm{\Gamma}$, as shown in Fig.~\hyperref[SHC-odd]{\ref{SHC-odd}(c)}. Therefore, errors in $\bm{\Gamma}$ do not significantly affect our conclusions, thus the assumption of strain-independent conductivity for $\mathrm{OsO}_2$ here is reasonable. Following this assumption, the $|\theta_{\text{AS}}|$ curve under different $\mathcal{E}_{\mathrm{ts}}$ is plotted after determining $\bm{\Gamma}$, shown as the red line in Fig.~\hyperref[SHC-odd]{\ref{SHC-odd}(d)}, which is compared with the maximum spin splitting near the Fermi surface without SOC: $|\mathrm{Splitting}|_{\mathrm{max}}$ (black line) and the intrinsic spin Hall angle (ISHA) $|\theta_{\text{IS}}|$ (blue line) associated with the relativistic CSHE. Discussions about CSHE in $\mathrm{OsO}_2$ will be detailed in Sec.~\ref{CSHE}. Specific values of $\bm{\Gamma}$, charge conductivity $\sigma_{xx}$, ASSC $|\sigma_{xy}^{z\text{,odd}}|$, and $|\theta_{\text{AS}}|$ are provided in Table~S3~\cite{supplement}.

\begin{table*}
  \caption{\label{tabeven}Calculated ISHC $\sigma_{xy}^{z}$ and corresponding ISHA $|\theta_{\text{IS}}|$ of $(001)$-oriented \ce{OsO2}, as well as $\sigma_{zx}^{z}$ and corresponding out-of-plane $|\theta_{\text{IS}}^{\perp }|$ of $(101)$-oriented \ce{OsO2} under different equibiaxial tensile strain $\mathcal{E}_{\mathrm{ts}}$.}
  \tabcolsep=0.50cm
  \begin{ruledtabular}
    \begin{tabular}{ccccc}
      $\mathcal{E}_{\mathrm{ts}}~(\%)$ & $\sigma_{xy}^{z}~[(\hbar/e)\mathrm{S/cm}]$    & $|\theta_{\text{IS}}|~(\%)$ & $\sigma_{zx}^{z}~[(\hbar/e)\mathrm{S/cm}]$   & $|\theta_{\text{IS}}^{\perp }|~(\%)$         \\ [3pt] \hline & \\[-2.0ex]
      $0$                              & $ 10.80 $                                      & $0.06$                     & $-      $                                    & $-   $                                       \\ [3pt]
      $1$                              & $-50.15 $                                      & $0.30$                     & $108.25 $                                    & $0.65$                                       \\ [3pt]
      $2$                              & $-141.37$                                      & $0.85$                     & $86.34  $                                    & $0.52$                                       \\ [3pt]
      $3$                              & $-153.17$                                      & $0.92$                     & $7.24   $                                    & $0.04$                                       \\ [3pt]
      $4$                              & $-134.09$                                      & $0.81$                     & $-87.53 $                                    & $0.53$                                       \\ [3pt]
      $5$                              & $-166.85$                                      & $1.00$                     & $-80.96 $                                    & $0.49$                                       \\ [3pt]
      $6$                              & $-280.55$                                      & $1.68$                     & $-311.87$                                    & $1.87$
      \end{tabular}
  \end{ruledtabular}
\end{table*}

It can be observed from Fig.~\hyperref[SHC-odd]{\ref{SHC-odd}(d)} that when SOC is neglected, the variation trend of $|\theta_{\text{AS}}|$ coincides with that of the spin splitting magnitude. When SOC is considered, the situation changes notably. $\mathrm{OsO}_2$ achieves the highest spin-charge conversion efficiency $|\theta_{\text{AS}}|\approx7\%$ at $\mathcal{E}_{\mathrm{ts}}=3\%$. This value significantly exceeds the ISHA $|\theta_{\text{IS}}|$ generated via CSHE in $\mathrm{OsO}_2$, and surpasses those of a series of $4d$ and $5d$ transition metals including Nb, Ta, Mo, Pd, and Pt~\cite{guo-prl2008,Pt2,Pt3}. It is noteworthy that when $\mathcal{E}_{\mathrm{ts}}=4\%$ and $5\%$, the $|\sigma_{xy}^{z\text{,odd}}|$ curve without SOC exhibits an overall enhancement compared to other curves, as shown by the purple and gold dashed lines in Fig.~\hyperref[SHC-odd]{\ref{SHC-odd}(a)}. This implies that under the same scattering rate $\bm{\Gamma}$, the introduction of SOC leads to a smaller ASSC, which may be related to the narrower spin splitting induced by SOC discussed in the previous section (Sec.~\ref{FermiSurface}). This effect can be directly observed in Figs.~\hyperref[025band]{\ref{025band}(d)} and~\hyperref[025band]{\ref{025band}(e)}. Meanwhile, in the absence of SOC, the electronic structure of $\mathrm{OsO}_2$ (Fig.~S1~\cite{supplement}) exhibits more electronic states near the Fermi level compared to the SOC-included case (Fig.~S3~\cite{supplement}), which leads to higher charge conductivity. As shown in Fig.~S7~\cite{supplement}, the $\sigma_{xx}$ without SOC is systematically shifted upward relative to the SOC-included $\sigma_{xx}$ [Fig.~\hyperref[SHC-odd]{\ref{SHC-odd}(b)}] within the same $\bm{\Gamma}$ range, manifested as larger $\bm{\Gamma}$ values corresponding to the experimental conductivity (Table~S3~\cite{supplement}). The complex interaction of these factors ultimately results in the maximum $|\theta_{\text{AS}}|$ at $\mathcal{E}_{\mathrm{ts}}=3\%$ in the presence of SOC [Fig.~\hyperref[SHC-odd]{\ref{SHC-odd}(d)}], demonstrating that the strong SOC in $\mathrm{OsO}_2$ is beneficial for its strain-induced ASSE. It is worth mentioning that we also investigate the magnetic moment of Ru variation with equibiaxial strain $\mathcal{E}_{\mathrm{s}}$ in $\mathrm{RuO}_2$ in the absence of Hubbard $U$ correction (Table~S4~\cite{supplement}). The results reveal that $\mathrm{RuO}_2$ exhibits no magnetism without strong Coulomb correlation, while minor equibiaxial strain can induce altermagnetism. This suggests the necessity for experimental characterization of the Hubbard $U$ parameter's value in $\mathrm{RuO}_2$. Weaker electronic correlations in $\mathrm{RuO}_2$ may explain the ongoing controversies regarding its magnetic properties.

\subsection{\label{CSHE}Relativistic CSHE in OsO$\mathbf{_2}$}

\begin{figure*}
  \includegraphics[width=1.0\textwidth]{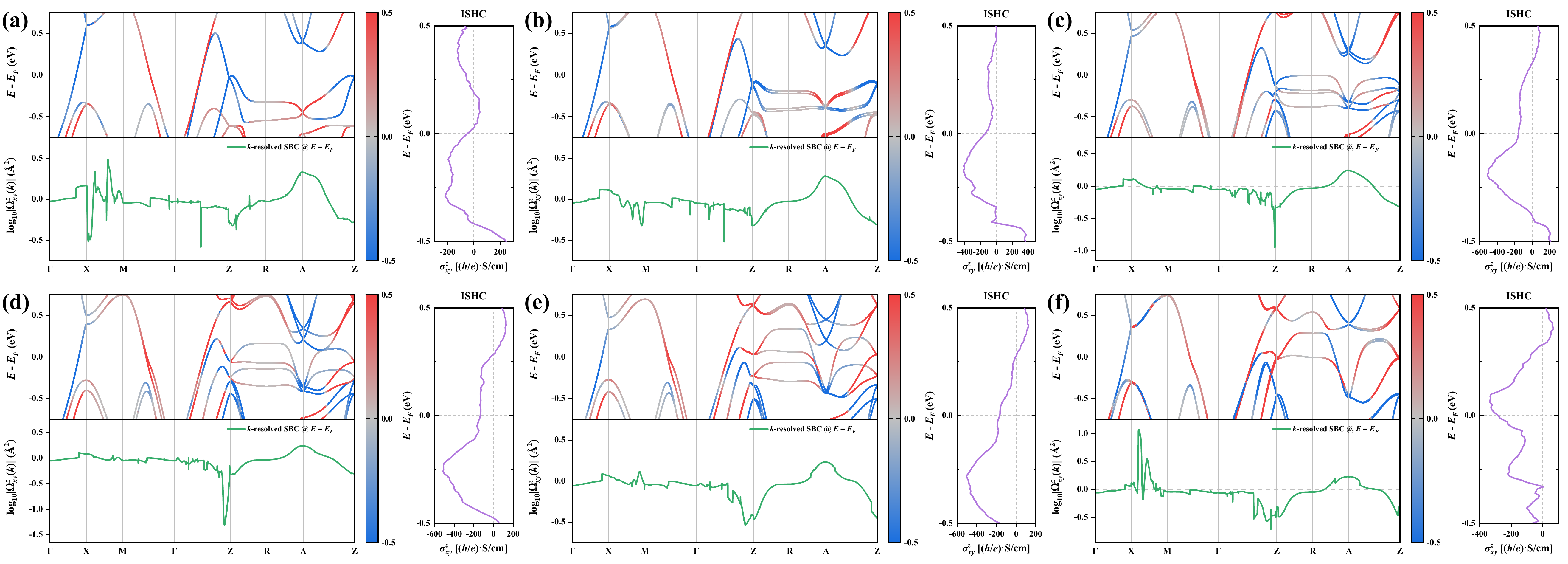}
  \caption{\label{ISHC}Band structures projected by SBC $\Omega_{xy}^{z}$ on a logarithmic scale using Eq.~(\ref{log}) and $k$-resolved SBC of $\mathrm{OsO}_2$ under $\mathcal{E}_{\mathrm{ts}}$ values of (a) $1\%$, (b) $2\%$, (c) $3\%$, (d) $5\%$, and (e) $6\%$. The ISHC $\sigma_{xy}^{z}$ as a function of Fermi energy is shown on the right.
  }
\end{figure*}

As a $5d$ heavy metal, Os exhibits strong SOC, which is expected to generate a large ISHC. This section will continue to investigate the effects of equibiaxial tensile strain $\mathcal{E}_{\mathrm{ts}}$ on the ISHC $\sigma_{xy}^{z}$ and SBC in $\mathrm{OsO}_2$, calculated using Eqs.~(\ref{eq:kubo_shc_berry}) and (\ref{eq:kubo_shc_berry_sum}), respectively. The calculated ISHC $\sigma_{xy}^{z}$ and the SBC $\Omega_{xy}^{z}$ resolved band structures as well as the $k$-resolved SBC of $\mathrm{OsO}_2$ with $\mathcal{E}_{\mathrm{ts}}$ in range of $1\%$ to $6\%$ are illustrated in Fig.~\ref{ISHC}. The corresponding results for $\mathcal{E}_{\mathrm{ts}}=0\%$ are shown in Fig.~S8~\cite{supplement}. The SBC is plotted by taking the logarithm of Eqs.~(\ref{eq:kubo_shc_berry}) and (\ref{eq:kubo_shc_berry_sum}) to more clearly show the rapid variation of SBC. This strategy was also employed in previous studies~\cite{log1,log2,log3}, and its definition is given by

\begin{equation}
\label{log}
\Omega'=
\begin{cases}
    \mathrm{sgn}(\Omega)\log_{10}{|\Omega|}\text{,}  &  |\Omega|> 10 \\
    \hfil \frac{\Omega}{10}\text{,} \hfil &  |\Omega|\leqslant   10
\end{cases}\text{,}
\end{equation}  
where sgn($\Omega$) means taking the sign of $\Omega$.

Table~\ref{tabeven} lists the values of ISHC $\sigma_{xy}^{z}$ at $E=E_{F}$ and ISHA $|\theta_{\text{IS}}|=|\sigma_{xy}^{z}/\sigma_{\text{exp}}|$ under different $\mathcal{E}_{\mathrm{ts}}$, where $\sigma_{\text{exp}}$ represents the experimental conductivity at room temperature, i.e., $\sim16666.67$~$\mathrm{S/cm}$~\cite{lattice1,OsO2}. Our calculated ISHC magnitude without strain (i.e., $\mathcal{E}_{\mathrm{ts}}=0\%$) is 10.80 $(\hbar/e)\mathrm{S/cm}$, which is close to previous computational values [9 $(\hbar/e)\mathrm{S/cm}$] in Ref.~\cite{OsO2-SHC}. From both Table~\ref{tabeven} and Fig.~\hyperref[SHC-odd]{\ref{SHC-odd}(d)}, it can be observed that the magnitude of ISHA $|\theta_{\text{IS}}|$ exhibits an almost positive correlation with $\mathcal{E}_{\mathrm{ts}}$, and it remains significantly lower than $|\theta_{\text{AS}}|$ when ASSE exists in $\mathrm{OsO}_2$ (i.e., $\mathcal{E}_{\mathrm{ts}}=2\%\sim5\%$). This further demonstrates the superiority of ASSE over CSHE and the potential breakthroughs brought by strain engineering.

\begin{figure}
  \centering
  \includegraphics[width=0.5\textwidth]{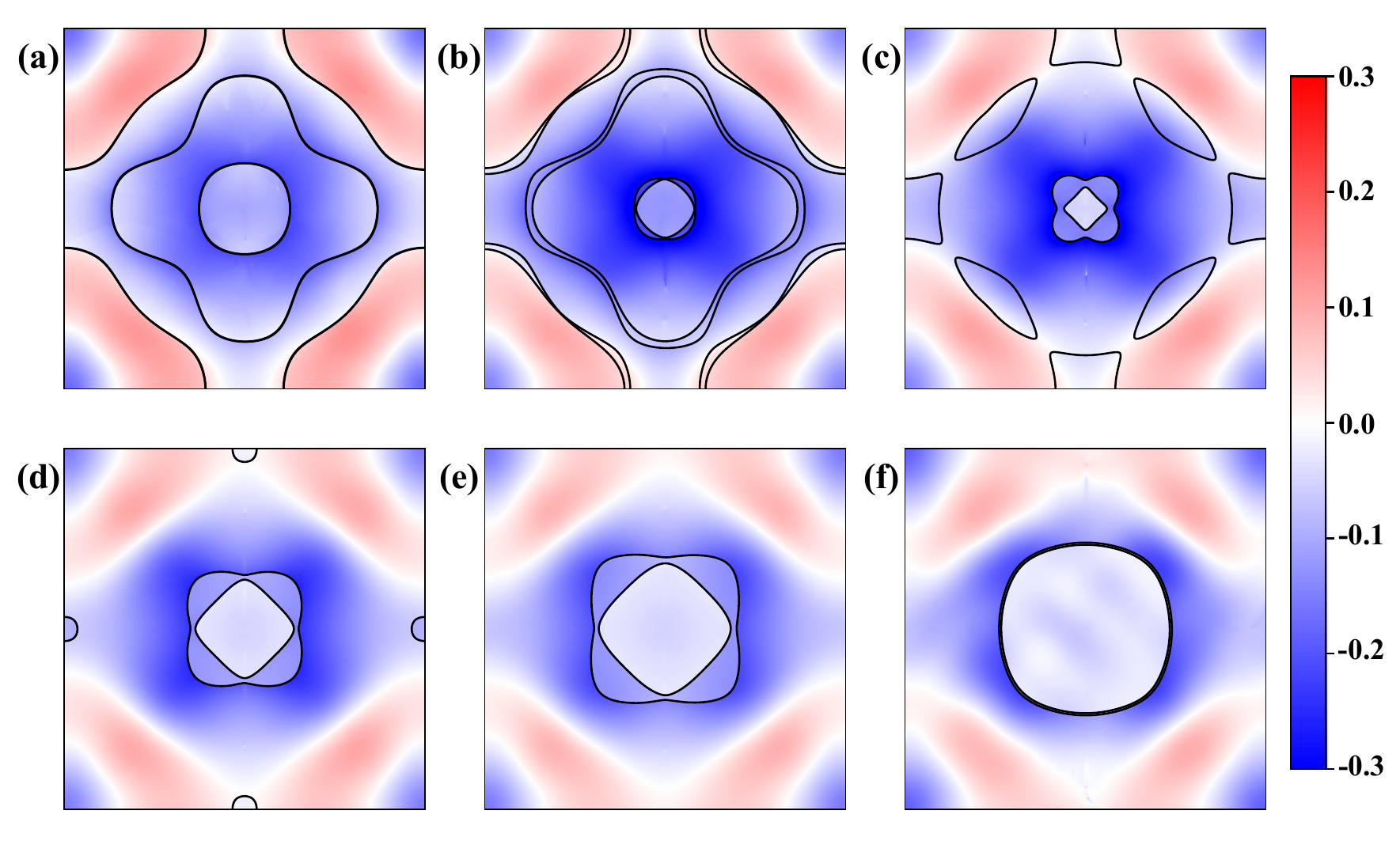}
  \caption{\label{SBC}The $k$-resolved SBC $\Omega_{xy}^{z}$ on a logarithmic scale calculated using Eq.~(\ref{log}) in a 2D BZ slice at the $k_{z}=\pi/2c$ plane of $\mathrm{OsO}_2$, under $\mathcal{E}_{\mathrm{ts}}$ values of (a) $1\%$, (b) $2\%$, (c) $3\%$, (d) $5\%$, and (e) $6\%$ at $E=E_{F}$. The black lines denote the intersections of the Fermi surface with the slices. The red and blue colors represent positive and negative SBC $\Omega_{xy}^{z}$ (in unit of \AA$^2$), respectively.
  }
\end{figure}

We know that ISHC is directly related to the SBC, which acts as a magnetic field in momentum space~\cite{CrXO}. Some sharp peaks caused by the bands crossing near the Fermi energy can be observed at specific $k$ paths in the $k$-resolved SBC, as illustrated in Fig.~\ref{ISHC}. It can be seen that strain has relatively minor effects on the band structure near the Fermi level of $\mathrm{OsO}_2$, which leads to similar $k$-resolved SBC patterns and comparable energy-dependent ISHC trends under different strains. We are particularly interested in how the alternating pattern at the $k_{z} = \pi/2c$ plane influences SBC, and therefore we calculated the projected $k$-resolved SBC on this slice, as shown in Fig.~\ref{SBC}. The overlapping regions of the two ellipses result in nearly vanishing SBC, while the elliptical tips enhance SBC, producing larger negative values. Beyond these features, the overall SBC remains largely unaffected. This observation further confirms the distinct physical origins between ASSE and CSHE.

As a supplement, we also offer the calculated $\mathcal{T}$-odd ASSC and $\mathcal{T}$-even ISHC tensors in units of $(\hbar/e)\mathrm{S/cm}$ for $\mathrm{OsO}_2$ with different growth orientations and charge current directions, as displayed in Tables~S5 and S6~\cite{supplement}. For $(001)$-oriented $\mathrm{OsO}_2$, only components $\sigma_{\alpha\beta}^{\gamma}$ with mutually orthogonal $\alpha$, $\beta$, $\gamma$ (i.e., the Levi-Civita tensor $\epsilon_{\alpha\beta\gamma} \ne 0$) are non-zero. For $(101)$-oriented $\mathrm{OsO}_2$, the coordinate system is transformed from that of a $(001)$-oriented $\mathrm{OsO}_2$ using a rotation matrix (counterclockwise rotation about the $y[010]$ axis by $\varphi$)
\renewcommand{\arraystretch}{0.75}
$$
\mathcal{D}=\begin{pmatrix}\cos\varphi & 0 & \sin\varphi \\0      & 1 &   0      \\-\sin\varphi & 0 & \cos\varphi\end{pmatrix}\text{,}
$$
where for the case of $(001)$-to-$(101)$-oriented rotation, $\varphi = \arctan\left(\frac{a}{c}\right)$. The tensors for $(101)$-oriented $\mathrm{OsO}_2$ are then obtained as follows~\cite{SST2} 

\begin{equation}
  \label{rot}
  \overset{(101)}{\sigma_{ij}^{k}}  = {\textstyle \sum_{l,m,n}}\mathcal{D}_{il}\mathcal{D}_{jm}\mathcal{D}_{kn}\overset{(001)}{\sigma_{lm}^{n}}.
\end{equation}

Similar to the case discussed in Ref.~\cite{SST2}, when an electric field is applied along the $[010]$ direction for $(101)$-oriented $\mathrm{OsO}_2$, the $\mathcal{T}$-even spin current flow in the $[001]$ direction exhibits both a component of flow in the vertical direction and a component of spin polarization out of plane, with amplitude 
$$
\overset{(001)}{\sigma_{zy}^{x}}\sin\varphi\cos\varphi\text{.} 
$$
Simultaneously, the spin current flow in the [100] direction also has a component of spin current flow in the vertical direction with a component of spin polarization out of plane, with amplitude 
$$
\overset{(001)}{\sigma_{xy}^{z}}\sin\varphi\cos\varphi\text{.}
$$
Thus, under the condition 
$$
-\overset{(001)}{\sigma_{zy}^{x}}=\overset{(001)}{\sigma_{zx}^{y}}\ne\overset{(001)}{\sigma_{xy}^{z}}\text{,}
$$
the $\mathcal{T}$-even mechanism can generate a non-zero net vertically-flowing spin current with out-of-plane spin polarization, characterized by
\begin{equation}
  \label{sigma}
  \overset{(001)}{\sigma_{zy}^{x}}\sin\varphi\cos\varphi + \overset{(001)}{\sigma_{xy}^{z}}\sin\varphi\cos\varphi = \overset{(101)}{\sigma_{zx}^{z}} = -\overset{(101)}{\sigma_{yx}^{y}} \text{,}
\end{equation}
which is crucial for field-free perpendicular magnetic switching. This particular spin current can also be generated through the $\mathcal{T}$-odd ASSE mechanism, with detailed calculation results shown in Table~S5~\cite{supplement}. At $\mathcal{E}_{\mathrm{ts}} = 3\%$, $|\sigma_{zx}^{z\text{,odd}}|$ reaches up to 646.96 $(\hbar/e)\mathrm{S/cm}$. However, unlike the case in the $\mathcal{T}$-even mechanism, in $\mathcal{T}$-odd ASSC tensors, only the components with spin polarization aligned with the direction of N\'eel vector $\hat{\mathcal{N}}$ can be generated in the absence of SOC~\cite{SST2,CrXO}. Actually, these components in $(101)$-oriented $\mathrm{OsO}_2$ will cancel each other due to $\hat{\mathcal{N}}$ along the $[001]$ and $[00\bar{1}]$ directions between different domains, i.e.,
$$
\overset{(101)^{\text{tot}}}{\sigma_{zx}^{z\text{,odd}}}=\overset{(101)^{\hat{\mathcal{N}}\uparrow}}{\sigma_{zx}^{z\text{,odd}}}+\overset{(101)^{\hat{\mathcal{N}}\downarrow}}{\sigma_{\bar{z}x}^{\bar{z}\text{,odd}}}\rightarrow 0\text{,}
$$
where $\hat{\mathcal{N}}\uparrow$ and $\hat{\mathcal{N}}\downarrow$ denote the N\'eel vector oriented along the $[001]$ and $[00\bar{1}]$ directions, respectively. Therefore, we focus exclusively on the $\mathcal{T}$-even mechanism, which is unconstrained by the N\'eel vector here. 

The calculated values of $\mathcal{T}$-even $\sigma_{zx}^{z}$ in $(101)$-oriented $\mathrm{OsO}_2$ under different $\mathcal{E}_{\mathrm{ts}}$ are displayed in Table~\ref{tabeven}, with full $\mathcal{T}$-even ISHC tensors detailed in Table~S6. For comparison, we also present $\mathcal{T}$-odd ASSC tensors with $\mathcal{E}_{\mathrm{ts}}$ ranging from $2\%\sim5\%$ in Table~S5~\cite{supplement}. It can be observed from Table~\ref{tabeven} that a $6\%$ equibiaxial tensile strain in $(001)$-oriented $\mathrm{OsO}_2$ is able to induce an out-of-plane ISHA $|\theta_{\text{IS}}^{\perp}|$ as high as $1.87\%$, far exceeding that in $\mathrm{RuO}_2$~\cite{SST2} and even surpassing the conventional ISHA in $(001)$-oriented $\mathrm{OsO}_2$. This can be simply achieved by rotating $\mathrm{OsO}_2$ to the $(101)$-oriented configuration. Through Table~S6, it is noted that as $\mathcal{E}_{\mathrm{ts}}$ increases, the value of $\sigma_{zy}^{x}$ transitions from positive to negative, while $\sigma_{xy}^{z}$ remains consistently negative. According to Eq. (\ref{sigma}), the value of $\sigma_{zx}^{z}$ in $(101)$-oriented $\mathrm{OsO}_2$ initially cancels out due to opposing signs as $\mathcal{E}_{\mathrm{ts}}$ increases, reaching a minimum at $\mathcal{E}_{\mathrm{ts}} = 3\%$. Subsequently, it enhances due to identical signs, ultimately attaining a maximum at $\mathcal{E}_{\mathrm{ts}} = 6\%$.

\section{Conclusion}

This work proposes strain engineering as a novel viable approach to induce ASSE, successfully achieving equibiaxial tensile strain $\mathcal{E}_{\mathrm{ts}}$-induced ASSE in $\mathrm{OsO}_2$. We observed the emergence of alternating patterns at the $k_{z} = \pi/2c$ plane and nonrelativistic spin splitting along specific high-symmetry paths, while evaluating the $\mathcal{T}$-odd ASSC $|\sigma_{xy}^{z\text{,odd}}|$ and spin-charge conversion ratio $|\theta_{\text{AS}}|$ generated by strain-induced ASSE in $\mathrm{OsO}_2$. Our findings reveal that $|\theta_{\text{AS}}|$ reaches a maximum of $\sim7\%$ under $\mathcal{E}_{\mathrm{ts}}=3\%$, and notably maintains $\sim5\%$ even without SOC---significantly surpassing the ISHA $|\theta_{\text{IS}}|$ generated by CSHE in $\mathrm{OsO}_2$. We also find that $|\theta_{\text{AS}}|$ exhibits a positive correlation with the magnitude of spin splitting, both showing an initial increase followed by a decrease as $\mathcal{E}_{\mathrm{ts}}$ grows, whereas $|\theta_{\text{IS}}|$ displays an almost monotonic increase with increasing $\mathcal{E}_{\mathrm{ts}}$. Calculations of spin Berry curvature (SBC) further demonstrate the distinct physical origins between ASSE and CSHE in $\mathrm{OsO}_2$. Additionally, we predict that when $\mathcal{E}_{\mathrm{ts}} = 6\%$, the $(101)$-oriented $\mathrm{OsO}_2$ will generate an out-of-plane ISHA $|\theta_{\text{IS}}^{\perp}|$ of $\sim1.87\%$, which makes it highly promising for applications in the field-free perpendicular magnetic switching. Finally, we find that strain can also induce altermagnetism in $\mathrm{RuO}_2$, which not only demonstrates the universal feasibility of strain engineering for altermagnetism generation but also provides a plausible explanation for the longstanding debate regarding the existence of magnetism in $\mathrm{RuO}_2$. Further calculations considering the on-site Coulomb interaction (details in \hyperref[sec:appendix]{Appendix} section) indicate that, under reasonable Hubbard $U$ correction, our proposed strain-induced ASSE remains effective, yielding a $|\theta_{\text{AS}}|$ of up to 16.36\% without SOC. Even under the assumption that $\mathrm{OsO}_2$ is intrinsically altermagnetic ($U=2.0$ eV case), where larger strains substantially complicate the behavior, a slight 1\% strain still improves the spin-charge conversion ratio (Table~S2~\cite{supplement}). This further highlights the robustness of strain engineering in enhancing the conversion efficiency. This work provides new insights for altermagnet discovery and establishes theoretical foundations for designing next-generation altermagnetic spintronic devices.

\begin{acknowledgments}
The authors would like to acknowledge the financial support from National Key Research and Development Program of China (Grant No. 2022YFA1602701), and National Natural Science Foundation of China (Grants No. 12204364, No. 12327806, and No. 12227806). Numerical calculation is supported by the High-Performance Computing Center of Wuhan University of Science and Technology.
\end{acknowledgments}

\section*{Data Availability}
The data that support the findings of this article are openly available~\cite{data}.

\appendix*
\section{Results with Hubbard \textit{U} correction}\label{sec:appendix}
\setcounter{figure}{0}
\setcounter{table}{0}
\renewcommand{\thefigure}{A\arabic{figure}}
\renewcommand{\thetable}{A\Roman{table}}

\begin{figure}
  \centering
  \includegraphics[width=0.45\textwidth]{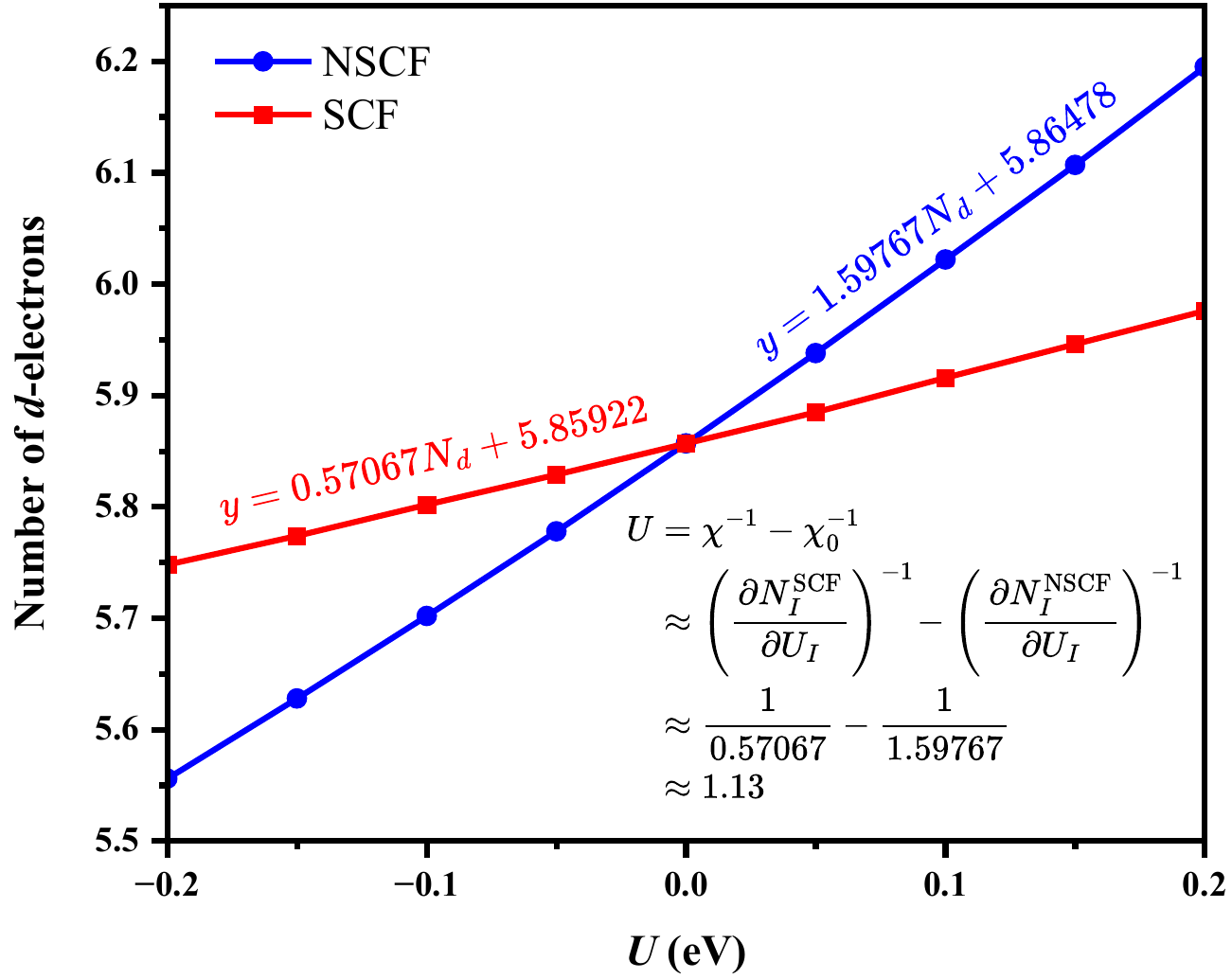}
  \caption{\label{FLLAU}Hubbard $U$ parameter determined by linear response approach.
  }
\end{figure}

As shown in Fig.~\ref{FLLAU}, the linear response approach~\cite{lla-U} indicates that $U = 1.13$ eV is appropriate for our system. Detailed results at $U=1.13$ eV are tabulated in Tables~S7 and S8~\cite{supplement}. Table~S7 presents the evolution of lattice constants, magnetic moment of Os atoms, and spin-resolved Fermi surface at the $k_{z}=\pi/2c$ with equibiaxial tensile strain $\mathcal{E}_{\mathrm{ts}}$. Table~S8 shows the dependence of scattering rate $\bm{\Gamma}$, charge conductivity $\sigma_{xx}$, ASSC $\sigma_{xy}^{z\text{,odd}}$, and spin-charge conversion ratio $|\theta_{\text{AS}}|$ on $\mathcal{E}_{\mathrm{ts}}$. For $U=1.13$ eV with $\mathcal{E}_{\mathrm{ts}}$ varying between 1\%--6\%, Fig.~S9 displays shapes of spin-resolved Fermi surfaces at different $k_{z}$ planes (the corresponding 3D Fermi surfaces are displayed in Fig.~S12), while Fig.~S10 shows the evolution of band structures with and without SOC. The case of $\mathcal{E}_{\mathrm{ts}}=0\%$ is displayed in Fig.~S13. Strain- and $\bm{\Gamma}$-dependent curves from Table~S8 are presented in Fig.~S11, with Fig.~S11(d) specifically showing $|\theta_{\text{AS}}|$ versus $\mathcal{E}_{\mathrm{ts}}$ at different $U$ values~\cite{supplement}.

Unlike the case without Hubbard $U$ correction, when $U=1.13$ eV, $\mathrm{OsO}_2$ exhibits a monotonic increase in both ASSC $\sigma_{xy}^{z\text{,odd}}$ and spin-charge conversion ratio $|\theta_{\text{AS}}|$ with increasing $\mathcal{E}_{\mathrm{ts}}$. At $\mathcal{E}_{\mathrm{ts}}=6\%$, $|\theta_{\text{AS}}|$ reaches 12.43\%, nearly twice the maximum value achieved without Hubbard $U$ correction. When SOC is excluded, $|\theta_{\text{AS}}|$ reaches up to 16.36\%. As clearly observed in Fig.~S10, the introduction of SOC reduces the spin splitting magnitude; this reduction is more pronounced than that observed without $U$ (Fig.~\ref{025band}), resulting in a larger $|\theta_{\text{AS}}|$ in the absence of SOC.

At $U=1.13$ eV, significant changes occur in the spin-resolved Fermi surface on the $k_{z}=\pi/2c$ plane. Compared to previous results (Fig.~\ref{FS}), it becomes more dispersed. At $\mathcal{E}_{\mathrm{ts}}=4\%$, no distinct alternating pattern is observable. As clearly seen in Fig.~S10(d), a bandgap opens at the Fermi level on the $k_{z}=\pi/2c$ plane, causing the disappearance of the alternating pattern, which shifts to other energy levels. Moreover, when $\mathcal{E}_{\mathrm{ts}}=4\%$ and without SOC, $\mathrm{OsO}_2$ becomes ferromagnetic [Fig.~S10(d)]. This situation is analogous to the case with $U=2.0$ eV, $\mathcal{E}_{\mathrm{ts}}=5\%$ (Table~S2~\cite{supplement}). However, the latter remains ferromagnetic when SOC is included, while the former becomes altermagnetic when SOC is considered.
Notably, when $|\theta_{\text{AS}}|$ reaches 6\%, the spin-resolved Fermi surface of $\mathrm{OsO}_2$ becomes distinctly asymmetric [Fig.~S9(f)]. Careful observation reveals that this asymmetry emerges as early as $|\theta_{\text{AS}}|=5\%$ [Fig.~S9(e) shows a larger red ellipse in the bottom-left corner compared to the top-right corner]. Additionally, under $U=1.13$ eV and zero strain, an alternating pattern persists in the spin-resolved Fermi surfaces despite a zero magnetic moment on Os atoms, consistent with observations made without $U$ correction. Currently, we cannot provide a definitive explanation for these phenomena, and exploring their underlying physical mechanisms lies beyond the scope of this work. Therefore, we refrain from further detailed discussion here. 

In summary, our calculations demonstrate that strain engineering can always enhance the ASSE-induced spin-charge conversion ratio in \ce{OsO2}, regardless of whether the on-site Coulomb interaction is considered or the material itself is altermagnetic. This robust and universal enhancement highlights the fundamental significance of our findings and underscores the broader applicability of strain engineering strategy for altermagnetic spintronics.

\bibliography{ref}

\end{document}